\documentclass{article}
\usepackage{bris}
\usepackage{cite}
\usepackage{epsfig}

\begin{document}
\begin{titlepage}{BRIS/HEP/2000--02}{July 2000}
\title{The Measurement of Azimuthal Asymmetries In Deep Inelastic
Scattering}

\author{N. H. Brook\\
(on behalf of the ZEUS Collaboration)}


\begin{abstract} 
The distribution of the azimuthal angle for the charged hadrons has been
studied in the hadronic centre-of-mass system for neutral current
deep inelastic positron-proton scattering at HERA.
Measurements of the dependence of the moments of this
distribution on the transverse momenta of the charged
hadrons are presented.
Asymmetries that can be unambiguously attributed to perturbative
QCD processes have been observed for the first time.
\end{abstract}
\end{titlepage}

\section{Introduction}
Semi-inclusive processes in deep inelastic scattering (DIS) are
of importance because they can be used to test the
perturbative Quantum Chromodynamic (QCD) description of hadron production
via parton fragmentation.
An observable of particular interest is the distribution
of the azimuthal angle, $\phi,$ (measured in the hadronic centre-of-mass
frame, HCM) between the lepton scattering
plane, defined by the incoming and outgoing lepton momenta,
and the hadron production plane, defined by the exchanged
virtual boson  and an outgoing hadron .

Asymmetries in the $\phi$ distribution, i.e. terms proportional to
$\cos \phi$ and $\cos 2\phi$ 
arise whenever a non-zero transverse momentum, in the HCM frame,
is present in the
scattering process.
Consequently both non-perturbative and perturbative QCD
effects~\cite{allazi,cahn,chay,gehrmann} give rise to these
asymmetries.
The azimuthal dependence  of parton production has the
form

\begin{equation}
    d\sigma/d\phi  = A + B \cos\phi + C \cos 2\phi .
\label{eq:azi}
\end{equation}

\noindent This form results from the polarisation of the exchanged
virtual boson.
The coefficients $B$ and $C$ depend on the helicities
of the final-state parton(s), partonic transverse momenta
and on colour coherence~\cite{chay}.
The
$\cos 2\phi$ term is expected from interference of amplitudes arising
from the $+1$ and $-1$
helicity components of the transversely-polarised part of the exchanged
boson,
whereas transverse/longitudinal interference gives rise to the
$\cos\phi$ term.
Terms in $\sin\phi$ and $\sin
2\phi$ are not expected for neutral current reactions. 

The asymmetry that results from the intrinsic momentum of a quark in the proton
is referred to as the non-perturbative asymmetry.
Since the intrinsic momentum is small, this asymmetry should
fall rapidly with increasing transverse momentum  of
the measured hadron and with
increasing $Q^2$~\cite{cahn},
where
$Q^2\equiv -q^2$
is the negative square of the four-momentum of the virtual exchanged boson.
In particular, the  term $C$ is small at high $Q^2.$ 
The transverse momenta arising from the fragmentation itself does not contribute
to the asymmetry but only smears the observed distribution~\cite{chay}.

In contrast, the asymmetry associated with leading-order terms in perturbative
QCD calculations, the perturbative asymmetry, is weakly dependent on $Q^2$ and 
persists at high transverse momenta.
The perturbative QCD contribution to
terms $B$ and $C$ is large 
at leading order (LO) in $\alpha_s.$ 
The $\cos 2\phi $ term can be unambiguously
attributed to perturbative QCD processes in the high $Q^2$ region under
investigation. 

The event kinematics of DIS are determined by 
$Q^2$, and one of the two 
Bjorken scaling variables $x=Q^2/2P\!\cdot\!q$
or $y = Q^2/xs,$ 
where $P$ is the four-momentum of the incoming proton
and $\sqrt s$ is the positron-proton centre-of-mass energy.
The kinematic region studied is $0.2 < y < 0.8$ and $ 0.01 < x < 0.1,$
corresponding to a $Q^2$ range $180 < Q^2 < 7220 {\rm\ GeV^2}.$

The form of Eq.~\ref{eq:azi} is expected to be maintained for single
particle production~\cite{chay,gehrmann}, since high-momentum hadrons are
produced close to the direction of the parton.
Note that in order to observe the $\cos \phi$ asymmetry, a selection
procedure is required which consistently associates leading
hadrons produced from
{\rm either} quarks {\rm or} gluons~\cite{chay,gehrmann}. 
Since the gluon fragmentation function is `softer'
than that of the quarks, the hadron-quark correlation can be enhanced by
selecting `leading' charged particles. This was accomplished
by cutting on the Lorentz-invariant variable
$z_h(=P\cdot p_h/P\cdot q)>0.2$, where 
$p_h$ is the track or particle four-vector.

\section{Results}
\begin{figure}[ht]
\centering
\begin{center}
\mbox{\epsfig{file=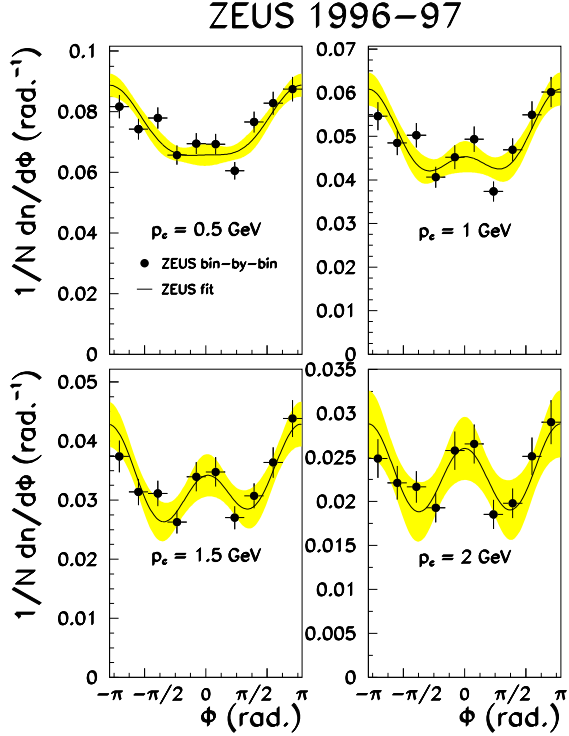,height=10.0cm}}
\end{center}
\caption{
The
differential $\phi$ distributions obtained
for four values of the transverse momentum
cut, $p_c,$ in the hadronic centre-of-mass frame
in the kinematic region $ 0.01 < x < 0.1$ and $ 0.2 < y < 0.8$
for charged hadrons with $0.2 < z_h < 1.0.$
The full line
(with accompanying statistical error band) is the
result obtained from the main analysis.
The data points were corrected using a bin-by-bin procedure (only
statistical errors are shown). 
}
\label{fig:phidist}
\end{figure}

Figure~\ref{fig:phidist} shows the 
differential $\phi$ distributions 
of charged hadrons 
for four values of their minimum  transverse momentum, $p_c,$ in the HCM.
The results obtained from the main analysis method~\cite{azipaper} 
(full line) and a bin-by-bin
correction method (points) are seen to agree.
At low $p_c,$ a clear $\cos \phi $ term is observed. As the value of 
$p_c$ is increased a $\cos 2\phi $ term becomes evident.  

\begin{figure}[ht]
\centering
\begin{center}
\mbox{\epsfig{file=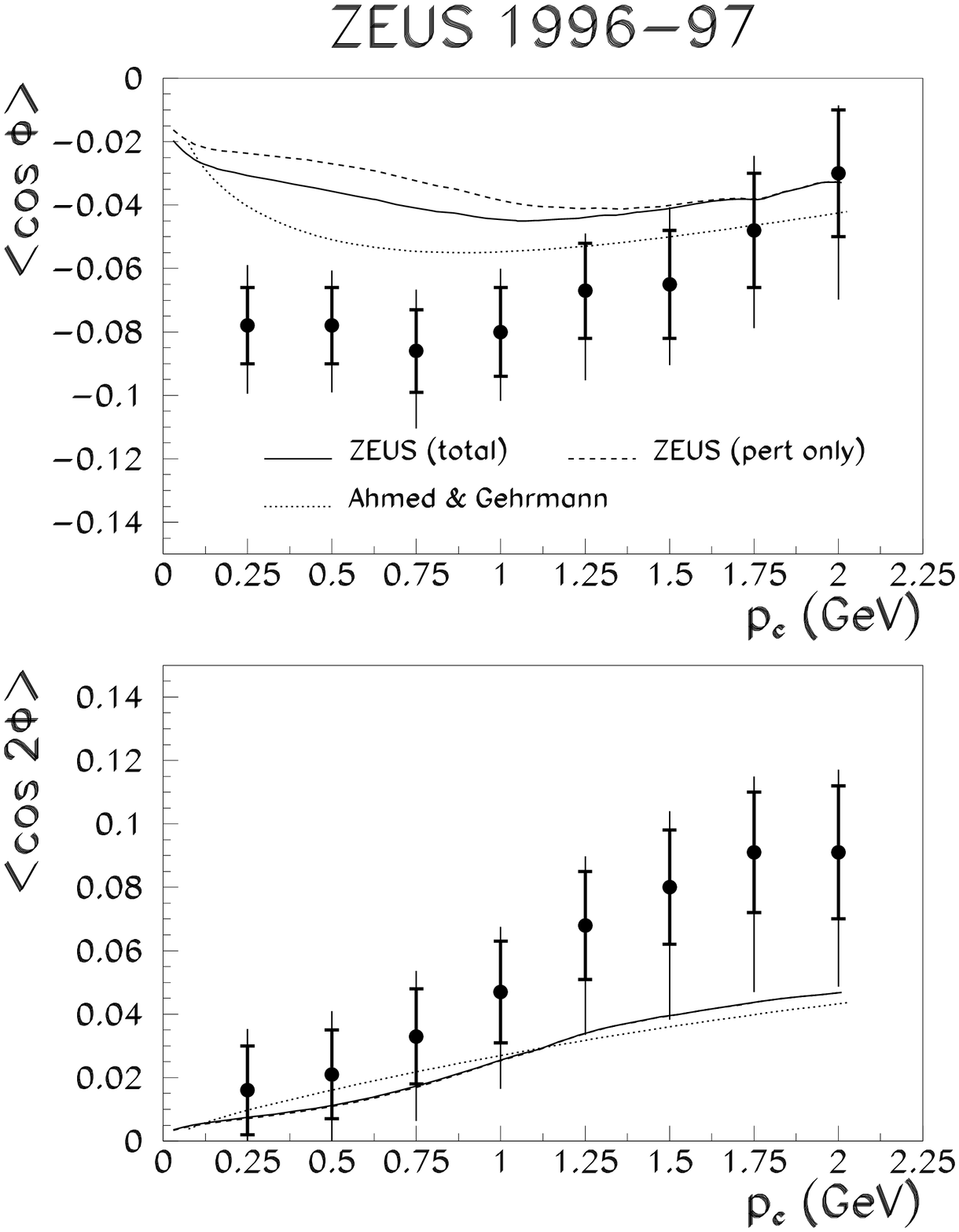,height=9.0cm}}
\end{center}
\caption{ The values of $\langle \cos \phi \rangle$
and $\langle \cos 2\phi \rangle$ are shown as a function of $p_c$ 
in the kinematic region $ 0.01 < x < 0.1$ and $ 0.2 < y < 0.8$
for charged hadrons with $0.2 < z_h < 1.0.$
The inner error bars represent the
statistical errors, the outer are statistical
and systematic errors added in quadrature.
The lines are the LO predictions from 
ZEUS with perturbative and non-perturbative 
contributions (full line),
ZEUS with the perturbative contribution only (dashed line)
and Ahmed \& Gehrmann  (dotted line -- see
text for discussion). For
the case of $\langle \cos 2\phi \rangle,$ the ZEUS total and
perturbative predictions are almost identical.}
\label{fig:mc_lo}
\end{figure}

Figure~\ref{fig:mc_lo} shows the moments
 $\langle \cos \phi \rangle$ and $\langle \cos 2\phi \rangle$
as a function of
$p_c.$ 
The $\sin\phi$ term
is consistent with zero independent of the value chosen for $p_c,$
in agreement with expectation.
The value 
of $\langle \cos \phi \rangle$ is negative and decreases in magnitude
as $p_c$ is increased. 
In contrast,
the value of $\langle \cos 2\phi \rangle$ is
positive and rises as $p_c$ is increased.
Figure~\ref{fig:mc_lo} also compares the data with two LO QCD calculations.
Both calculations were made with $Q$ as the
appropriate scale, with the Binnewies et al. LO fragmentation
function~\cite{binnewies} and with the CTEQ4
LO proton parton densities~\cite{CTEQ}.

The calculation from ZEUS (based
on the calculation of Chay et al.~\cite{chay}) 
includes an estimation of the non-perturbative
contribution, from intrinsic $k_T$ and hadronisation $p_T,$
and integrates over the whole kinematic range.
The results of Ahmed~\& Gehrmann are purely at
leading order in $\alpha_s$ and are evaluated 
at the mean values $\langle x \rangle=0.022$ and $\langle Q^2 \rangle=750\
{\rm GeV^2}$ of the data.
The different implementations
account for the observed difference in the two predictions;
using $\langle x \rangle$ and $\langle Q^2 \rangle$ in the ZEUS
perturbative calculation leads to agreement with the 
Ahmed~\& Gehrmann calculation.

\section{Conclusions}
The azimuthal asymmetries in the deep inelastic electroproduction of single
particles have been measured at HERA 
in the HCM.
For hadrons produced at large 
transverse momenta, the measured value for  $\langle \cos\phi \rangle$
is negative and is in agreement with QCD
predictions.  
The moment  $\langle \cos 2\phi \rangle$ 
has been measured here for the first time and is
non-zero and positive.   It increases as a function of the minimum
particle transverse momentum, as expected from QCD.
Since the non-perturbative contribution to
$\langle \cos 2\phi\rangle $ is predicted to be
negligible,
this measurement provides clear
evidence for a perturbative QCD contribution to the azimuthal asymmetry.

\end{document}